# Radiation-hard ASICs for optical data transmission in the ATLAS pixel detector

K.K. Gan, K.E. Arms, M. Johnson, H. Kagan, R. Kass, C. Rush, S. Smith,
R. Ter-Antonian, M.M. Zoeller

Department of Physics, The Ohio State University, Columbus, OH 43210, USA

A. Ciliox, M. Holder, M. Ziolkowski

Fachbereich Physik, Universitaet Siegen, 57068 Siegen, Germany

We have developed two radiation-hard ASICs for optical data transmission in the ATLAS pixel detector at the LHC at CERN: a driver chip for a Vertical Cavity Surface Emitting Laser (VCSEL) diode for 80 Mbit/s data transmission from the detector, and a Bi-Phase Mark decoder chip to recover the control data and 40 MHz clock received optically by a PIN diode. We have successfully implemented both ASICs in 0.25 µm CMOS technology using enclosed layout transistors and guard rings for increased radiation hardness. We present results from prototype circuits and from irradiation studies with 24 GeV protons up to 57 Mrad (1.9 x $10^{15}$ p/cm$^2$).

## 1. INTRODUCTION

The ATLAS pixel detector [1] consists of two barrel layers and two forward and backward disks which provide at least two space point measurements. The pixel sensors are read out by front-end electronics which is controlled by the Module Control Chip (MCC). The low voltage differential signal (LVDS) from the MCC chip is converted by the VCSEL Driver Chip (VDC) into a single-ended signal appropriate to drive a Vertical Cavity Surface Emitting Laser (VCSEL). The optical signal from the VCSEL is transmitted to the Readout Device (ROD) via a fiber.

The 40 MHz beam crossing clock from the ROD, bi-phase mark (BPM) encoded with the data (command) signal to control the pixel detector, is transmitted via a fiber to a PIN diode. This BPM encoded signal is decoded using a Digital Opto-Receiver Integrated Circuit (DORIC). The clock and data signals recovered by the DORIC are in LVDS form for interfacing with the MCC chip.

The ATLAS pixel optical link contains 448 VDC and 360 DORIC chips with each chip having four channels. The chips will be mounted on 180 chip carrier boards (opto-boards). The optical link circuitry will be exposed to a maximum total fluence of $10^{15}$ 1-MeV $n_{eq}$/cm$^2$ during ten years of operation at the LHC. In this paper we describe the development of the radiation-hard VDC and DORIC circuits for use in the ATLAS pixel detector's optical link.

## 2. VDC CIRCUIT

The VDC is used to convert an LVDS input signal into a single-ended signal appropriate to drive a VCSEL in a common cathode array. The output current of the VDC is to be variable between 0 and 20 mA through an external control current, with a standing current (dim current) of ~1 mA to improve the switching speed of the VCSEL. The rise and fall times of the VCSEL driver current are required to be less than 1 ns and the duty cycle of the VDC output signal should be (50 ± 4)%. In order to minimize the power supply noise on the opto-board, the VDC should also have constant current consumption independent of whether the VCSEL is in the bright (on) or dim (off) state.

Figure 1 shows a block diagram of the VDC circuit. An LVDS receiver converts the differential input into a single-ended signal. The differential driver controls the current flow from the positive power supply into the anode of the VCSEL. The VDC circuit is therefore compatible with a common cathode VCSEL array. An externally controlled current, $I_{set}$, determines the amplitude of the VCSEL current (bright minus dim current), while an externally controlled voltage, tunepad, determines the dim current. The differential driver contains a dummy driver circuit which in the VCSEL dim state draws an identical amount of current from the positive power supply as is flowing through the VCSEL in the bright state. This enables the VDC to have constant current consumption.

circuits are sensitive to power supply noise, we utilize two identical preamp channels: a signal channel and a noise cancellation channel. The signal channel receives and amplifies the input signal from the anode of the PIN diode, plus any noise picked up by the circuit. The noise cancellation channel amplifies noise similar to that picked up by the signal channel. This noise is then subtracted from the signal channel in the differential gain stage. To optimise the noise subtraction, the input load of the noise cancellation channel should be matched to the input load of the signal channel (PIN capacitance) via a dummy capacitance.

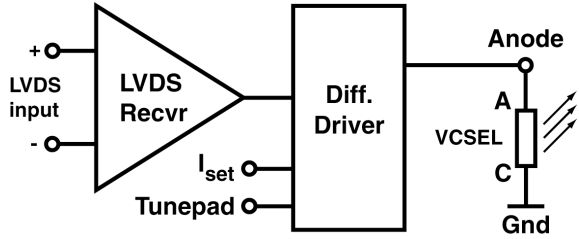

Figure 1: Block diagram of the VDC circuit.

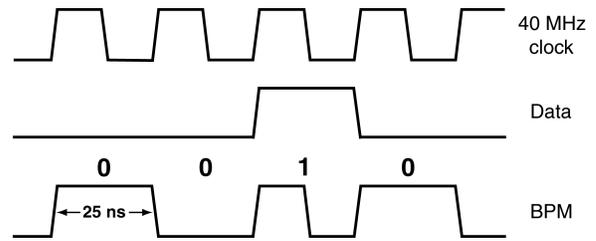

Figure 2: Example of a BPM encoded signal.

## 3. DORIC CIRCUIT

The function of the DORIC is to decode BPM encoded clock and data signals received by a PIN diode. Figure 2 shows an example of a BPM encoded signal. It is derived from the 40 MHz beam crossing clock by sending only transitions corresponding to clock leading edges. In the absence of data bits (logic level 1), this results simply in a 20 MHz clock. Any data bit in the data (command) stream is encoded as an extra transition at the clock trailing edge.

The amplitude of the current from the PIN diode is expected to be in the range of 40 to 1000 µA. The 40 MHz clock recovered by the DORIC is required to have a duty cycle of (50 ± 4)% with a total timing error of less than 1 ns. The bit error rate of the DORIC circuit is required to be less than $10^{-11}$ at end of life.

Figure 3 shows a block diagram of the DORIC circuit. In order to keep the PIN bias voltage (up to 10 V) off the DORIC chip, we employ a single-ended preamp circuit to amplify the current produced by the PIN diode. Since single-ended preamp

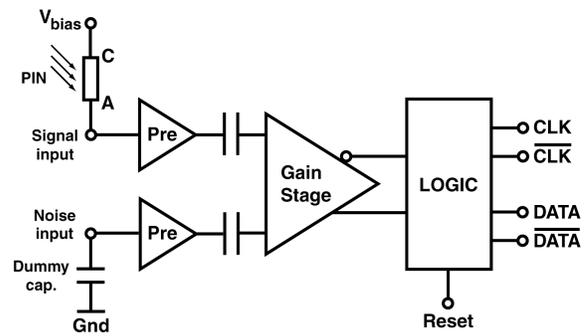

Figure 3: Block diagram of the DORIC circuit.

To ensure that the DORIC logic locks onto the correct recovered clock frequency, it needs to be trained over a certain time period. The specifications currently allow for a 25 µs initialization period, during which a 20 MHz clock is sent to the DORIC chip. The 40 MHz recovered clock is the input to a delay locked loop that adjusts the internal delays until a 50% duty cycle is reached. After having locked into the correct duty cycle, the clock recovery circuit is blind to any extra transitions near the middle of the 25 ns intervals, and will continue to decode the clock correctly even in the presence of data bits.

The data recovery circuit uses each input transition to latch the state of the recovered clock just prior to the current transition into a flip-flop (Figure 4). For no data sent, the recovered clock is always in a low state prior to each input transition. When a data bit is present however, the recovered clock is in a high state just prior to the input transition. The thus decoded data bit is then stretched to make it more stable during full clock cycle (see dotted and solid data bit in Figure 4).

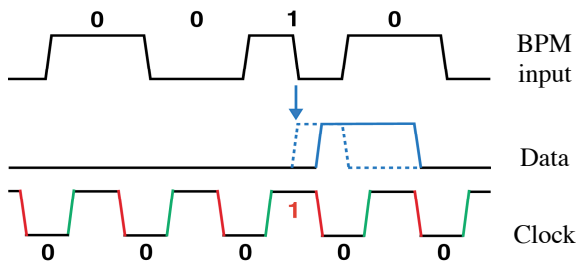

Figure 4: Data recovery procedure of DORIC logic.

## 4. VDC AND DORIC DESIGN HISTORY

The pixel detector design of the VDC and DORIC takes advantage of the development work for similar circuits [2] used by the outer detector, the SemiConductor Tracker (SCT). Both SCT chips attain radiation-tolerance by using bipolar integrated circuits (AMS 0.8 μm BICMOS) and running with high currents in the transistors at 4 V nominal supply voltage. These chips are therefore not applicable for the higher radiation dosage and lower power budget requirements of the pixel detector.

We originally implemented the VDC and DORIC circuits in radiation-hard DMILL 0.8 μm technology with a nominal supply voltage of 3.2 V. Using three submissions from summer 1999 to May 2001 we developed circuits in this technology that met the specifications. However, an irradiation study of the DMILL circuits in April 2001 with 24 GeV protons at CERN showed severe degradation of circuit performance. We concluded that the DMILL technology did not meet the radiation hardness requirement of the ATLAS pixel detector.

We have therefore migrated the VDC and DORIC designs to standard deep submicron (0.25 μm) CMOS technology which has a nominal supply voltage of 2.5 V. Employing enclosed layout transistors and guard rings [3], this technology promises to be very radiation hard. Four deep submicron prototype runs of the VDC and DORIC circuits have been received between summer of 2001 and summer 2002. Below we summarize the results achieved in the submissions.

## 5. RESULTS FROM IBM 0.25 MICRON SUBMISSIONS

Over the course of the four submissions, the VDC's total current consumption has been reduced and the current consumption between the bright and dim states of the VCSEL diode has been made more constant. Since the final VCSEL array will have a common cathode, the latest submission of the VDC circuit was made compatible with the array.

In Figure 5 we show the VCSEL current generated by the VDC from the various submissions as a function of the external control current $I_{set}$. The measurements are consistent with the expectations from the simulations. The current was measured with 10 Ω in series with the VCSEL, which partially explains the early turnover of the current at high values of $I_{set}$. We note that the VDC circuit from the third submission (compatible with a common anode VCSEL array) achieved 20 mA VCSEL current. The output current of the latest VDC design therefore still needs to be improved. We also observe that the dim current of this latest VDC circuits needs to be increased to ~1 mA. Further, we observe fairly balanced current consumption on the latest VDC circuits, and find that the duty cycle of the output signal is within specifications. The rise and fall times of the output signals are in the range of 1.0 to 1.4 ns over the operating range of the circuit. Since the VDC needs to be compatible with common cathode VCSELs, the switching of the current into the VCSEL is now performed by pFETs instead of faster nFETs. Clearly, the speed of the pFET switching circuitry still needs to be improved.

In the DORIC circuit, a feedback loop with a large time constant was added to fully cancel the offsets at its differential gain stage which vary from chip to chip. This also results in less clock jitter, since the preamp output signal is more stable. The gain of the preamp was reduced to make it less sensitive to noise pickup, and the flicker and white noise at the preamp output were also reduced. In order to reduce the coupling of digital signals into the preamp even further, an improved grounding scheme has been implemented, and the analog section has been well separated from the digital section of the circuit. Guard rings around the analog and digital sections have also been implemented.

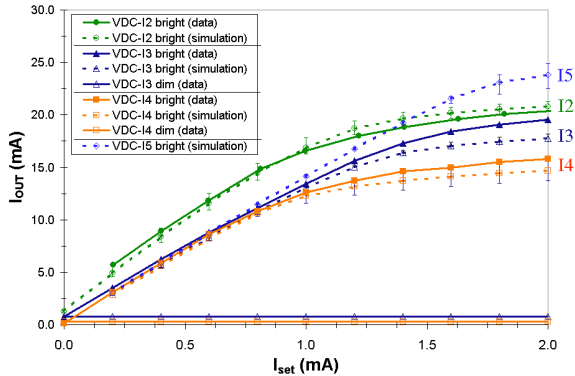

Figure 5: Comparison of the measured and predicted VCSEL drive current vs. $I_{set}$ for various versions of VDCs.

In the third deep submicron submission, the differential preamp was replaced by a single-ended preamp. This allows for the PIN diode to be biased directly, keeping the up to 10 V high bias voltage off the DORIC chip. The chip was able to decode control data and clock correctly down to PIN diode currents of ~25 µA, meeting the requirement of PIN current thresholds below 40 µA. This demonstrates that our single-ended preamp design matches the performance of the differential preamp design from the previous submission.

In the latest submission, the single-ended preamp was designed to be compatible with a common anode PIN array which are biased directly with a common negative supply. The preamp was further optimised for faster rise time at its output to ensure proper operation at small input amplitudes. The digital circuitry was also improved in several ways. The delay locked loop circuitry was optimised to ensure a duty cycle of the recovered clock closer to 50%, the timing of internal signals was optimised, and a reset circuit was added which allows slow and controlled recovery of the delay control circuit.

All of the improvements implemented in the DORIC-I4 circuit were successful. We observe however that the preamp circuit is very sensitive to the matching of the strays at the inputs to the signal and dummy channels. In order to test the DORIC-I4 circuit with a common cathode PIN array (which is the only version available), we shift the operating point of each preamp input by adding a 4.7 kΩ resistor to ground. With this setup, the DORIC-I4 performs adequately for PIN current amplitudes up to ~600 µA. In this operating region the jitter of the recovered 40 MHz clock is below 1 ns with a duty cycle close to 50% and a clock period close to 25 ns, which meets the requirements. On fully populated opto-boards, we routinely achieve PIN current thresholds for no bit errors of ~15 µA with the four-channel DORIC-I4.

There will be one more submission since the DORIC preamp needs to be redesigned to be compatible with a common cathode PIN array.

## 6. IRRADIATION STUDIES

We have irradiated 13 DORICs and 13 VDCs from the first 0.25 µm submission with 24 GeV protons at CERN in September 2001 up to a dosage of 50 Mrad (1.7 x $10^{15}$ p/cm$^2$). We observed no degradation in the amplitude and clock duty cycle of the output of the VDC. For the DORIC, the PIN current threshold for no bit errors remained constant. We received the irradiated chips one month after the irradiation and found that one chip had a much higher threshold. It was unclear whether the observed degradation of this chip was due to radiation or mishandling.

In August 2002, we irradiated VDC and DORIC circuits from the fourth deep submicron submission in the same proton beam at CERN. In a so-called cold box setup, we performed electrical testing of single channel VDC and DORIC circuits. For the 10 tested DORIC circuits we observed that the PIN current thresholds remained constant at ~12 µA up to the total dose of 57 Mrad. We found the bright and dim VCSEL currents for the eight tested VDC circuits to be constant throughout the irradiation. The duty cycle of the output signals, shown in Figure 6, increased by ~2% after 57 Mrad, which is acceptable.

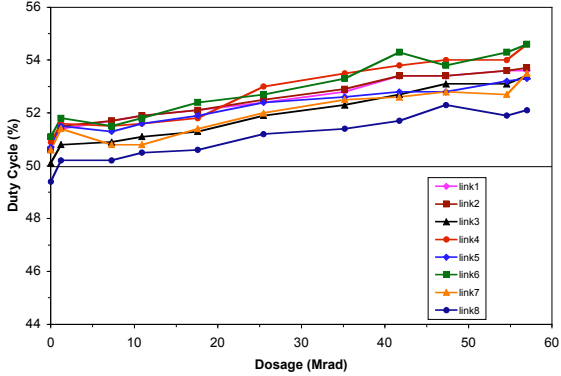

Figure 6: Clock duty cycle of VDC-I4 vs. dosage.

We have compared the performance of the VDC-I4 chips a month after irradiation to their performance before the irradiation. We observe no significant degradation of circuit performance on VCSEL drive current as shown in Figure 7. There is also no significant change in the rise/fall time, clock duty cycle and current consumption.

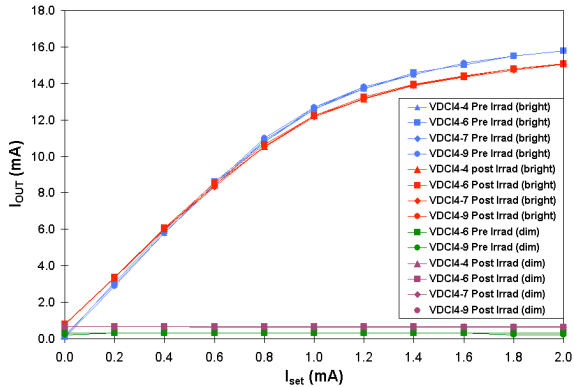

Figure 7: VCSEL drive current of VDC-I4 before and after irradiation.

In Figure 8 we show the PIN current thresholds for no bit errors, measured on one of the opto-boards during the irradiation study. We observe that the PIN current thresholds for no bit errors are ~10–15 mA and remain constant up to the total dose of 20 Mrad. One of the links developed a low upper PIN current threshold, ~40 mA, for no bit errors near the end of the irradiation. However, a measurement performed a month after the irradiation yields a much higher upper threshold, ~640 mA. We have also monitored the optical power returned from the opto-board to the control room over the course of the irradiation, as displayed in Figure 9. We were able to observe a general trend during the irradiation: after an irradiation period, the returned optical power usually decreased; after the VCSEL annealing period, the optical power usually increased, as expected. The variations from this general rule are most likely due to inconsistencies in the quality of the optical connections that had to be made for each measurement of the optical power. In future irradiation tests, we plan to improve on the quality and repeatability of the optical connections. We observed that the returned optical power from the opto-board decreased by as much as 60% after the total dose of 23 Mrad due to insufficient time for adequate annealing. We expect the links to recover most of the power after further annealing.

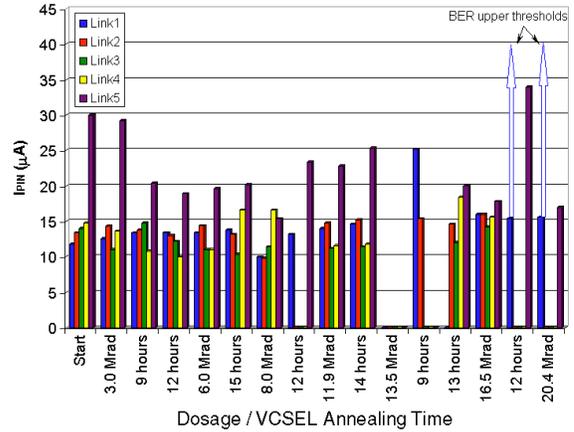

Figure 8: Opto-board PIN current thresholds for no bit errors vs. dosage. The VCSELs were annealed with 20 mA current during the indicated periods. There is no measurement at 13.5 Mrad. The thresholds could not be measured in a few occasions due to intermittent connection problems.

An additional opto-board, populated with single channel DORIC and VDC chips, was irradiated up to a total dose of 32 Mrad. The results from this board are similar to the results obtained with the two opto-boards described above.

## 7. SUMMARY

We have developed VDC and DORIC chips in deep submicron (0.25 µm) technology using enclosed layout transistors and guard rings for improved radiation hardness. The prototype circuits meet all the requirements for operation in the ATLAS pixel optical link and further appear to be sufficiently radiation hard for ten years of operation at the LHC.

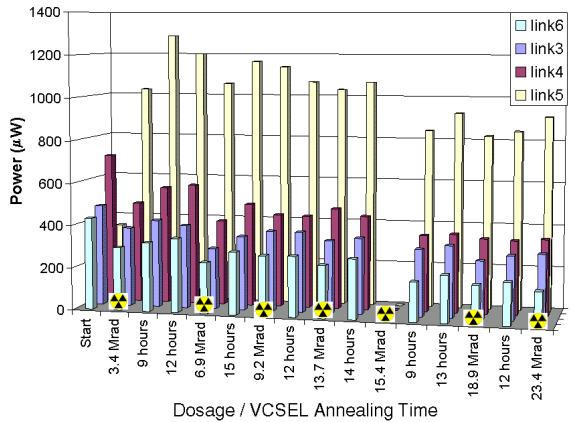

Figure 9: Opto-board optical power vs. dosage. The VCSELs were annealed with 20 mA current during the indicated periods. There is no measurement at 15.4 Mrad.

## ACKNOWLEDGEMENTS

This work was supported in part by the U.S. Department of Energy under contract No. DE-FG-02-91ER-40690.